\documentclass[journal,10pt]{IEEEtran}

\usepackage{stfloats}
\usepackage[english]{babel}
\usepackage{blindtext}
\usepackage[english]{babel}
\usepackage{blindtext}
\usepackage{booktabs} 
\usepackage{amsopn}
\usepackage{graphicx}
\usepackage{subcaption}
\usepackage{mathtools}
\usepackage{amsmath}
\usepackage{multirow}
\usepackage[noend]{algpseudocode}
\usepackage{epstopdf}
\usepackage{amsthm}
\usepackage{mathtools}
\usepackage{pgfplots}
\usepackage{caption}
\usepackage{amsmath, pgfplots, graphicx, siunitx}
\usepackage{amsmath,pgfplots,gincltex}
\usepackage{multirow}
\usepackage{siunitx}
\usepackage{amsmath, pgfplots, graphicx, siunitx, tikzscale}

\usepackage{algorithm}
\usepackage{algpseudocode}

\usepackage{graphicx}
\usepackage{caption}
\usepackage{acronym}
\usepackage{textcomp}

\makeatletter
\newcommand{\specificthanks}[1]{\@fnsymbol{#1}}
\makeatother

\newcommand{\hide}[1] {}

\definecolor{mypink1}{rgb}{0.858, 0.188, 0.478}

\usepackage[font=normalsize]{caption}
\usepackage{url}
\newcommand{\ie}{\emph{i.e.}\xspace}
\newcommand{\eg}{\emph{e.g.}\xspace}

\newcommand{\etc}{etc.\xspace}
\newcommand{\etal}{et al.\xspace}
\begin{document}
	
	\title{RACE: \textit{R}einforced  Cooperative \textit{A}utonomous Vehicle \textit{C}ollision Avoidanc\textit{E}}
	
	\author{Yali~Yuan\textsuperscript{\specificthanks{2}}
		\and , Robert~Tasik\textsuperscript{\specificthanks{2}}
		\and ,  Sripriya Srikant~Adhatarao,
		Yachao~Yuan,
		Zheli~Liu~\IEEEmembership{Member,~IEEE,}
		and~Xiaoming~Fu,~\IEEEmembership{Senior Member,~IEEE}
		\thanks{Yali Yuan, Robert Tasik, Sripriya Srikant Adhatarao and Xiaoming Fu are in Computer Networking Group, Institute of Computer Science, University of Goettingen, 37077 Goettingen, Germany. Emails: yali.yuan, fu, sripriya-srikant.adhatarao@cs.uni-goettingen.de, robert.tasik@stud.uni-goettingen.de.}
		\thanks{Yachao Yuan is in Smart Mobility Research Group, Institute of Information Systems, University of Goettingen, 37075 Goettingen, Germany. Email:yachao.yuan@uni-goettingen.de.}
		\thanks{Zheli Liu is in college of computer and control engineering, Nankai University, Tianjin, China. Email: liuzheli@nankai.edu.cn.}
		
		\thanks{\specificthanks{2} Yali Yuan (corresponding author) and Robert Tasik have equal contributions and are both first authors.}
	}
	
	{}
	
	\maketitle
	
	\begin{abstract}
		With the rapid development of autonomous driving, collision avoidance has attracted attention from both academia and industry. 
		Many collision avoidance strategies have emerged in recent years, but the dynamic and complex nature of driving environment poses a challenge to develop robust collision avoidance algorithms. 
		Therefore, in this paper, we propose a decentralized framework named \emph{\textbf{RACE}: \textbf{R}einforced Cooperative \textbf{A}utonomous Vehicle \textbf{C}ollision Avoidanc\textbf{E}}. 
		Leveraging a hierarchical architecture we develop an algorithm named \emph{Co-DDPG} to efficiently train autonomous vehicles. 
		Through a security abiding channel, the autonomous vehicles distribute their driving policies. We use the relative distances obtained by the opponent sensors to build the VANET instead of locations, which ensures the vehicle's location privacy.
		With a leader-follower architecture and parameter distribution, RACE accelerates the learning of optimal policies and efficiently utilizes the remaining resources. 
		We implement the RACE framework in the widely used TORCS simulator and conduct various experiments to measure the
		performance of RACE. 
		Evaluations show that RACE quickly learns optimal driving policies and effectively avoids collisions. 
		Moreover, RACE also scales smoothly with varying number of participating vehicles. 
		We further compared RACE with existing autonomous driving systems and show that RACE outperforms them by experiencing 65\% less collisions in the training process and exhibits improved performance under varying vehicle density.
	\end{abstract}
	\IEEEpeerreviewmaketitle
	\section{Introduction}
	The 2018 global road traffic accident report from the World Health Organization~\cite{global} revealed that an estimated 1.35 million people died worldwide due to road accidents. 
	Even though the rate of death relative to the world's population remains a constant, the United Nations' goal to realize 50\% reduction in road accidents by 2020 remains a distant dream.
	Moreover, there is no observed reduction in road accidents in low income countries.
	The cost to endure traffic accidents per year is approximated to 518 billion US dollars \cite{cost_of_accidents,gnp_cost}; approximately 1\% of the gross national product (GNP) in low income countries.
	Most of the road or traffic accidents are caused by collisions \cite{rolison2018factors}.
	Substituting the error-prone human driving with autonomous vehicles using artificial intelligence has the potential to circumvent the collision problems.
	Nevertheless, developing efficient computing models for robust autonomous vehicles that replicate the desired driving behaviour to avoid the collisions is still an ongoing research challenge \cite{wang2018networking}.
	
	Widespread research in the field of collision avoidance in autonomous driving has led to numerous solutions ranging from control theoretic formalization \cite{liniger2019noncooperative} and optimal control methods \cite{deng2019cooperative} to potential field-based and rule-based techniques~\cite{statheros2008autonomous}. 
	Recent advancements in the field of machine learning with imitation learning and deep reinforcement learning have facilitated improved data-driven approaches to minimize collisions. 
	For example, in \cite{chen2018cognitive}, Min \etal predicted collisions using neural networks with optimized genetic algorithms and back propagation and in \cite{chen2017decentralized}, Chen \etal explored collision avoidance using deep reinforcement learning with multiple agents.
	However, these approaches cannot effectively scale to accommodate the increasing road safety requirements. 
	Further, many collision avoidance algorithms use precise information (\eg, location) of the agents\footnote{Please note we use the terms agent and vehicle interchangeably.} \cite{jiang2012location,Su2009Power} to avoid collisions and thereby compromise the privacy of participating agents during autonomous driving.
	
	Moreover, we observed that, autonomous vehicles learn their surrounding environment and optimal driving strategies through a large amount of onboard sensors, such as lidar, radar and acceleration sensors. However, because of the limited communication range of the vehicles, the collected information from one vehicle is not sufficient to fulfill large-scale road safety requirements and improve the collision performance efficiently.  
	Therefore cooperation among the vehicles is a promising solution to overcome the above-mentioned shortcoming.
	In cooperated vehicle communication, the vehicles can communicate with each other. Each vehicle can learn their neighbor's surrounding environment. 
	As a result, the cooperated vehicles communication enables the vehicles to incorporate essential environmental information and thereby improve their autonomous driving behaviors.
	However, Kaushik \etal~\cite{kaushik2018parameter} show that, cooperated vehicle communication strategy faces the challenges of enforcing vehicles within lanes and prevent unnecessary over-takings. 
	They proposed parameter sharing to resolve these issues, but failed to take the scalability of their solution into account since with increasing number of agents the collisions increased exponentially.
	Some recent works \cite{riegger2016centralized,alrifaee2014centralized,loayza2017centralized} also proposed centralized solutions, where a server deployed at the central location computes efficient collision avoidance strategies for all participating agents.
	The centralized server maintains the comprehensive knowledge about all the agents as they send their current state information to the server and wait for the most intelligent driving strategy from the server.
	However, such centralized algorithms cannot effectively scale under heavy workload, especially when many agents are simultaneously requesting for driving instructions. 
	Moreover, such systems heavily depend on the availability of a secure and private communication network between the agents and the server, which is hard to guarantee in an agile driving environment \cite{ning2019attacker}. 
	Therefore, further research is required to provide optimal autonomous driving solutions that satisfy the desired driving requirements.
	especially efficient collision avoidance solutions which also fulfill the privacy needs of the users \cite{luo2018wireless,luo2019localization}.
	
	Based on the above discussion, we propose \textbf{RACE: \textit{R}einforced Cooperative \textit{A}utonomous Vehicle \textit{C}ollision Avoidanc\textit{E}}, a robust and efficient autonomous driving framework to overcome the above mentioned shortcomings. 
	In RACE, we provide a hierarchical multi-agent deep reinforcement learning model that exploits the well-known Deep Deterministic Policy Gradients (DDPG) \cite{ddpg} algorithm to effectively avoid collisions during autonomous driving. 
	We leverage the cooperated vehilce communication model, and develop a private and dynamic ad-hoc vehicular network algorithm based on VANETs, to build hierarchical vehicular networks with leading and following agents. 
	Using the distribution of parameters from the leading agent to the following agents, we foster a resource efficient real-time deep reinforcement collision avoidance algorithm. 
	
	RACE is designed to meet real-world autonomous driving design goals and hence it can rapidly adapt to the changes in the driving environments, such as sparsity or density of vehicles at intersections, and reduce the number of collisions. 
	During learning, RACE enforces varying levels of stringent penalties when collisions or potential situations, which may lead to collisions, occur; in order to build a robust learning environment. 
	Employing the TORCS \cite{wymann2015torcs} simulator, we implement RACE in a non-deterministic real-world car driving scenario in a racing environment. 
	With evaluations using the TORCS simulator, we show that RACE effectively learns from the environment and its neighboring agents.
	As the agents learn, we observe an exponential decrease in the number of collisions and a stable increase in rewards during learning.
	We compare RACE with existing baselines from Pinxin \etal \cite{long2018towards} and show that RACE outperforms existing works and dramatically reduces the collisions by 65\% and improves resource utilization by exploiting the multi-level agents. 
	Interestingly, we also observed that due to the varying levels of penalties designed in RACE, with sufficient learning, the vehicles chose to go off track in a potential collision scenario rather than collide with other vehicles or nearby objects to minimize the risk and liability.
	Thus, RACE provides a robust learning environment with the proposed Co-DDPG algorithm, which converges quickly and distributes the best parameters among the participating agents. 
	The main contributions in this work include:
	\begin{itemize}
		\item A novel hierarchical collision avoidance framework named \emph{RACE} to administer an efficient autonomous driving environment. 
		Through leveraging the deep reinforcement learning and the cooperated vehicle communication, RACE provides an efficient autonomous driving system that achieves the desired driving goals and ensures the location privacy of vehicles. 
		\item A dynamic hierarchical ad-hoc mobile network creation algorithm based on VANETs to create and maintain vehicular networks during autonomous driving. 
		The network is used to communicate necessary information to the participating agents in order to ensure smooth driving experience.
		\item An improved deep reinforcement learning algorithm named \emph{Co-DDPG} to efficiently learn the driving patterns. 
		Employing parameter distribution, the learned driving patterns are communicated with neighbors to utilize the resources efficiently and share the learning experiences among agents. 
		\item Extensive evaluations using TORCS simulator to demonstrate the benefits of RACE in comparison to existing solutions in an autonomous driving environment.
	\end{itemize} 
	\section{System Design}
\begin{figure*}[!t]
		\centering
		\begin{minipage}{.40\textwidth}
			\centering
			\hspace{-2mm}
			\includegraphics[width=0.8\linewidth]{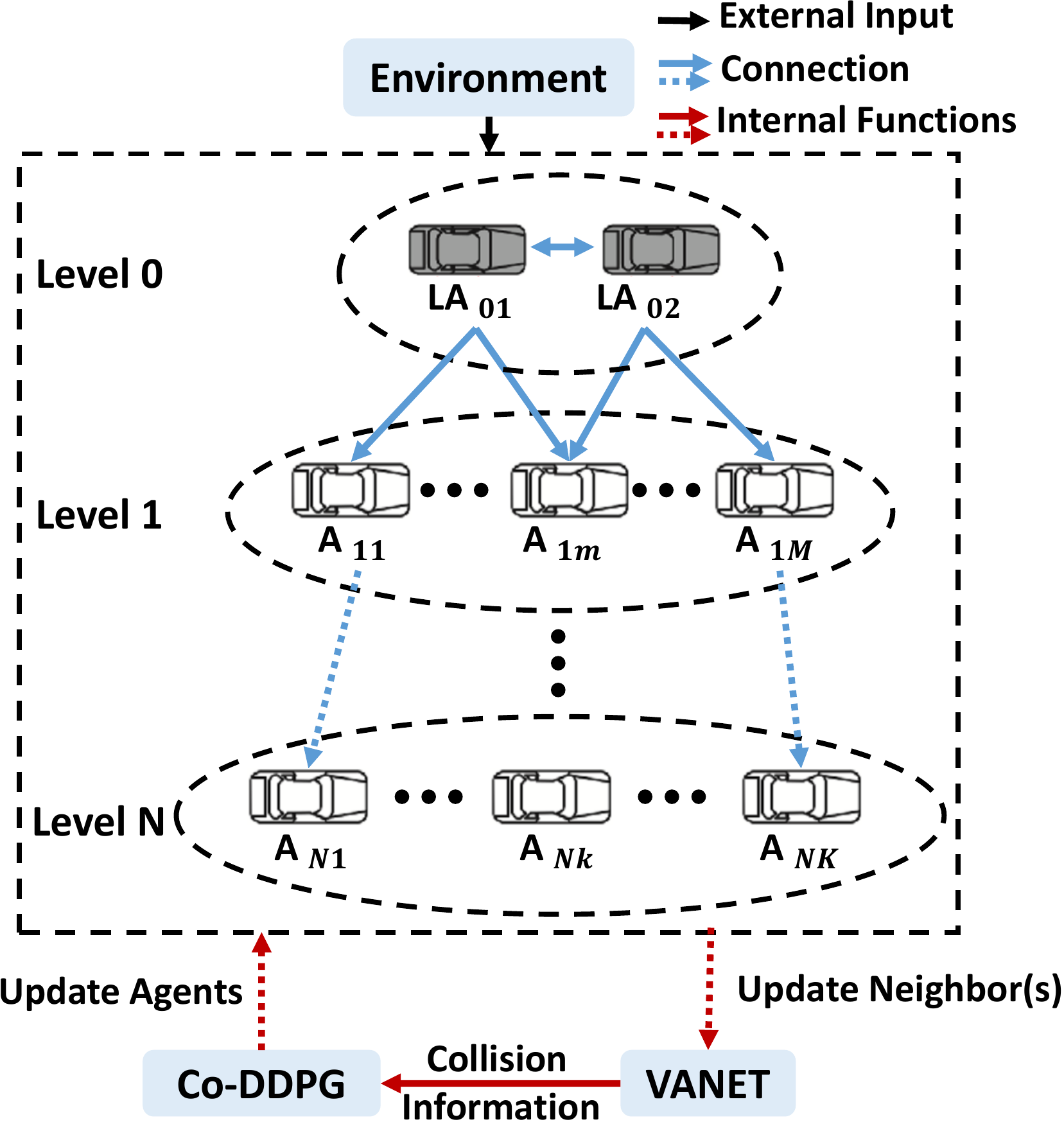}\\
			\vspace{2mm}
			(a) RACE Framework.
		\end{minipage}
		\hspace{1.6mm}
		\begin{minipage}{.40\textwidth}
			\centering
			\vspace{14mm}
			\includegraphics[width=0.65\linewidth]{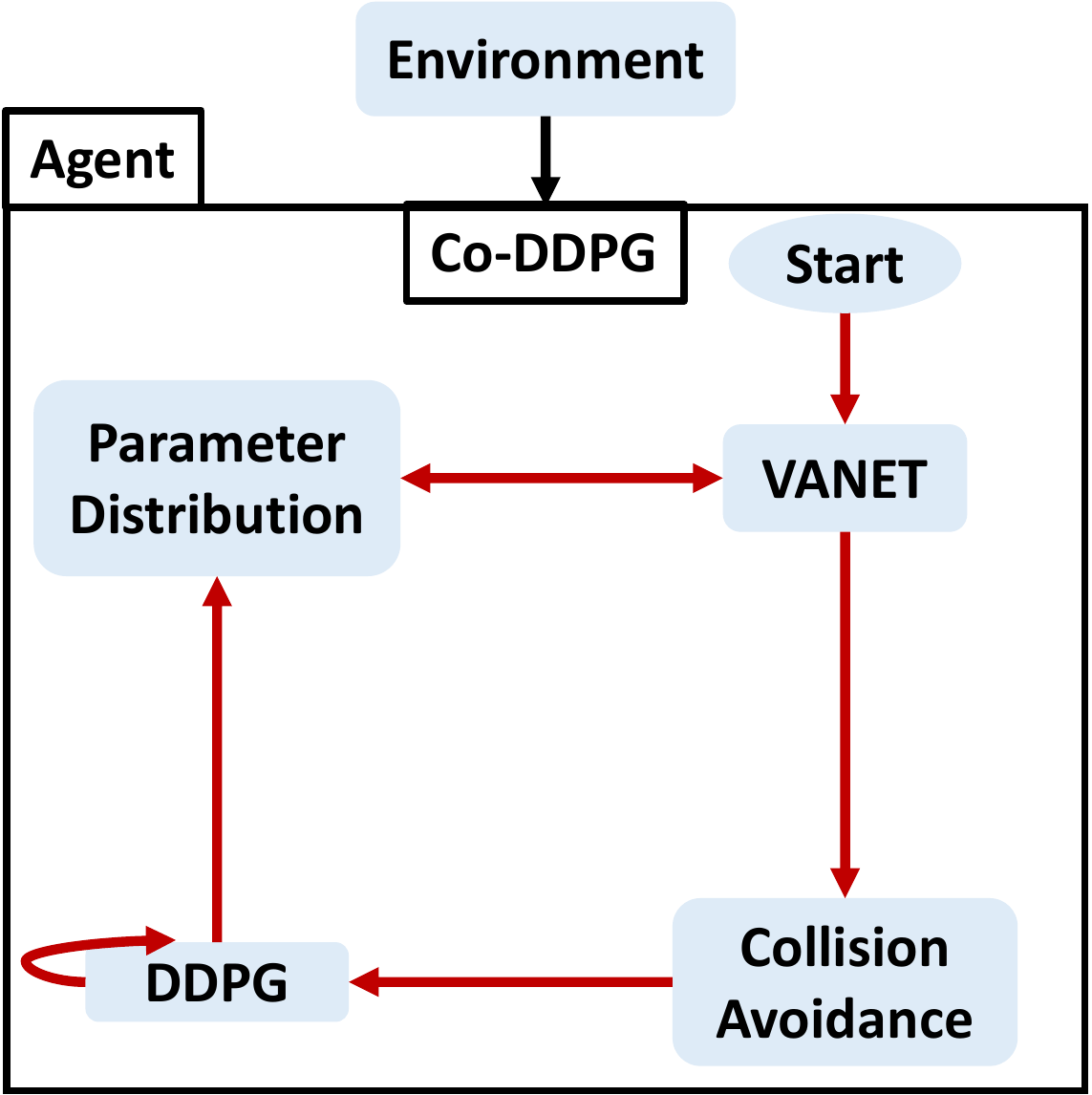}\\
			\vspace{2mm}
			(b) System Overview of Agent.
		\end{minipage}%
		\vspace{-1mm}
		\caption{RACE Architecture.}
		\label{fi:framework}
		\vspace{-6mm}
	\end{figure*}

	In this section, we elaborate the design goals of autonomous driving and the rationale behind them. 
	This is followed by a detailed description of the RACE framework and a use case example. 
	Figure \ref{fi:framework} shows an overview of the proposed RACE framework along with its internal system components to realize the design goals of an autonomous driving system.

	\subsection{Design goals}
	An autonomous driving system should fulfill the following design goals:
	\begin{itemize}
		\item Risk mitigation: It is crucial for an autonomous driving system to avoid any potential collision and damage in order to offer safe transportation with minimal risk. 
		\item Robustness: An autonomous driving system should be able to overcome technical difficulties arising during driving and opt for safer alternatives to avoid impending danger.
		\item Uncertainty: An autonomous driving system should be able to account for unpredictable and unexpected complications arising from the surrounding environment and neighboring vehicles. 
		\item Resource utilization: An autonomous driving system should optimally consume the available resources, such as computational resources and energy. 
		This is crucial to ensure an efficient and safe driving experience. 
		Moreover, many resources are often depleted at a faster rate during driving and are often difficult to replenish in some scenarios. 
		\item Efficiency: In autonomous driving, with increasing number of agents the complexity of machine learning models also increases. 
		The learning algorithm should be able to scale easily and handle the learning complexity efficiently in order to generate accurate driving actions in real-time. 
		\item Location Privacy: As the information is distributed in an open access environment in VANET, drivers' location privacy is critical and has to be satisfied to provide a reliable system.
	\end{itemize}
	\vspace{-4mm}
	\subsection{Architectural Components}
	The proposed RACE framework, which satisfies the above mentioned design goals is illustrated with an example scenario in Figure~\ref{fi:framework} along with an overview of the system in every autonomous vehicle. 
	The framework in Figure~\ref{fi:framework}(a) is composed of $N$ layers organized hierarchically where, $LA_{ij}$ represents the leading agent $LA_j$ in the $i$th layer. Correspondingly, $A_{(i+1)m}$ is the following agent $A_{m}$ in the $(i+1)$th layer of the leading agent $LA_{ij}$. The agents at level 0 are always considered as leading agents, whereas the agents at subsequent layers can take up the role of both leading and following agents based on their level in the hierarchy.
	The main components of this framework are described as follows:
	
	\begin{itemize}
		
		\item Environment: It represents the collective information regarding various road conditions gathered by the on board sensors in the autonomous vehicles \eg, distance towards the track edge, speed, angle, \etc 
		We build dynamic VANETs to facilitate the agents to communicate with each other and update their environmental conditions. 
		
		\item VANET: This component creates a dynamic vehicular ad-hoc network, \ie VANET to rapidly find any nearby vehicle(s) and create a network for communication and efficient driving. 
		Essentially, every agent identifies its neighboring autonomous vehicles and adds them to its local list of neighbors to which it is already connected.
		This module is mainly responsible for opening, leaving and closing the connections of an agent with its neighbors in a VANET.
		
		\item Hierarchical Agent Network: This is a submodule of the VANET. 
		It is created to dynamically allocate agents with their respective roles in the VANET. 
		The roles are designated based on varying levels of their available resources to ensure efficient operation. 
		This hierarchical network is composed of many layers with various agents found on different layers. 
		However, we can broadly classify the agents into two categories, namely leading agents and following agents. 
		A leading agent represents an autonomous vehicle at a layer with maximum available resources in comparison to its following agents located at the lower layers. 
		Every agent stores its corresponding following agents in a list similar to the neighbors list used during the VANET creation. 
		
		\item Collision Avoidance: This component tries to minimize the collisions during driving by allotting penalties for undesired driving behaviours such as crashing or driving too closely with other participating vehicles and rewards for good driving behaviours like keeping on the road, avoiding collisions and minimizing the arrival time. 
		
		\item Co-DDPG: Cooperative Deep Deterministic Policy Gradients (Co-DDPG) algorithm is an enhanced version of the standard DDPG algorithm designed in RACE. 
		Along with autonomous learning, it also facilitates collision avoidance with neighboring agents during deep reinforcement learning within a VANET.
		
		\item Parameter Distribution: The parameter distribution enables leading agents in RACE to accumulate deep learning parameters through Co-DDPG, such as weights, biases and rewards and distribute them to their following agents.

	\end{itemize}

	\begin{figure}[!t]
		\centering
		\begin{minipage}{.44\textwidth}
			\centering
			\includegraphics[width=0.95\linewidth]{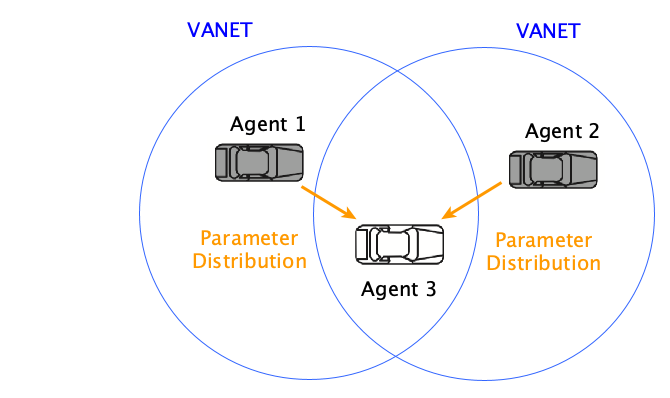}
			\vspace{-3mm}
			\caption{Parameter distribution.}
			\label{fig:parameterdist}
		\end{minipage}
		\vspace{-6mm}
	\end{figure}
	
	\subsection{Parameter Distribution}
	\label{subsec:parameterdist}
	One major component in the system design of Co-DPPG is the \emph{Parameter Distribution} module. 
	In RACE, the parameter distribution is facilitated by the creation of VANETs, which ensure that all participating vehicles can communicate with each other. 
	The crux of parameter distribution is based on comparing the \textit{average reward} accumulated by each agent for every action executed during autonomous driving.
	As this reward in reinforcement learning represents the incentives for good driving behaviour, it enforces and enriches the desired learning process. 
	Therefore, we consider the average reward as a suitable metric to design the parameter distribution in RACE. 
	Please note that in RACE, the following agents can choose to either learn or wait for distributed parameters from their leading agent(s) based on their available resources.
	If a following agent chooses to wait, it updates its parameters with the distributed parameters from a leading agent with highest average reward.
	Otherwise, a following agent compares its average reward based on its learned parameters with that of the leading agent. 
	The following agent will update its parameters with the distributed parameters, if the leading agent's average reward is higher than its current average reward.
	Otherwise, the following agent will forward its parameters to the leading agent to ensure that all agents learn the best policy.
	For instance in Figure~\ref{fig:parameterdist}, Agent 1 and Agent 2 distribute two different set of parameters to their following agent \ie, Agent~3. 
	Based on the average rewards associated with these two parameter sets, Agent 3 will select the parameters with the highest possible reward and incorporates it into its learning model. 
	
	\subsection{Design}
	We acknowledge that it is challenging to implement a multi-agent autonomous driving learning environment. 
	This is mainly due to the increasing complexity and uncertainty of the participating agents and their responses to other agents and the varying environmental conditions. 
	Each agent explores the learning environment with an aim to find the best policy under current circumstances and thereby increase the complexity of its learning. 
	Subsequently, this also influences the learning complexity of other participating agents. 
	With increasing number of agents, the resulting resource consumption also increases and thus, an inefficient learning environment will negatively impact the desired goals of autonomous driving.
	Therefore, we introduce an efficient decentralized learning algorithm named Co-DDPG (see Section \ref{sec-coddpg}) based on deep reinforcement learning and parameter distribution in RACE to ensure adequate learning in a multi-agent environment that satisfies the desired learning goals. 
	
	The RACE framework shown in Figure \ref{fi:framework}(a) employs a hierarchical agent network, where each agent runs the Co-DDPG algorithm with its supporting system components shown in Figure \ref{fi:framework}(b). 
	The Co-DDPG algorithm enables reinforcement learning parameters such as weights and biases to be shared among all the connected agents in the VANET. 
	The initial parameters are distributed by a primary leading agent that instantiates the vehicular network. 
	For instance in Figure~\ref{fig:parameterdist}, an initial network was created by the leading Agent 1 and leading Agent 2 and thus, they distribute the initial learning parameters to their subsequent following agents \ie, Agent 3. 
	Each agent's model parameters are updated whenever there exists a better trained model in the network. 
	With such a learning behaviour, all agents in the VANET benefit from distributed and learned experiences with each other. 
	Some agents may receive parameters from two or more leading agents 
	\eg, Agent 3 in Figure \ref{fig:parameterdist} received two parameter distributions, one each from Agent 1 and Agent 2. In such a scenario, the agents will select the best parameters from the received choices. 
	This ensures that, in RACE always the best collision avoidance parameters are shared among the agents to maintain a safe distance between the participating vehicles and to stay on the road by avoiding the edges of the road. 
	Specifically, every agent in RACE is organized in to a hierarchical leader-follower architecture. 
	The leading agents learn from the current environment and distribute their learned parameters, specifically the weights and bias, to their following agents. 
	The following agents can choose to either learn from the environment similar to their leading agent(s) or wait and download the parameters from their leading agent(s) based on their available resources. 
	If the following agents choose to learn from the environment, they compare their learned parameters with the received parameters from their leading agents and store the best policy. 
	If the following agent learns the best policy, then the following agent communicates the best policy to its leading agent. 
	This ensures that all agents in the current environment learn the best possible driving behaviours.
	
	Essentially, with $N$ number of layers organized hierarchically, the set of all layers is represented as $\mathcal{L}=\{L_{0}, L_{1}, \cdots, L_{i}, \cdots, L_{N}\}$. 
	If the agents in layer $L_{i}$ are designated as leading agents, then the agents in the following layer $L_{i+1}$ are considered to be the followers of the agents in the layer $L_{i}$. 
	The number of agents in each layer is denoted as $M$, where $M\in [0, N]$ and $M$, $N$ are integers. 
	The set of all agents in layer $i$ is given as $\mathcal{V}=\{A_{i1}, \cdots, A_{im}, \cdots, A_{iM}\}$. 
	Initially, two agents in the top most layer are selected as the leading agents. 
	The agents run the Co-DDPG (see Section \ref{sec-coddpg}) learning algorithm to find the optimal driving strategy and how to interact with other neighboring vehicles. 
	The following agents in the communication range of the leading agent estimate their remaining available resources and choose to either learn by running their Co-DDPG algorithm or  wait for their leading agents to distribute the parameters. 
	After the leading agents distribute their trained parameters, the following agents select the parameters from the leading agents 
	and updates their current Co-DDPG parameters with the received parameters.
	Since we optimally organize the agents based on their available resources into leader or follower and further provide the followers with an option to wait in order to conserve their remaining resources, RACE guarantees optimal utilization of the available resources. 
	\subsection{Use case Example}
	Let us consider a car driving on a road as an example use case to describe the operation of a system implemented using the proposed RACE framework. 
	Initially, the road is empty, and at time $T_i$, where $i\in [0, N]$, a single agent, Agent 1, enters the road and starts driving. 
	Since there are no neighboring agents to connect and synchronize the driving, Agent 1 learns its best driving behaviour primarily from Co-DDPG.
	At time $T_{i+1}$, a second agent, Agent 2, enters the road and soon comes in the connection range of Agent 1. 
	Agent 1 and Agent 2 initialize the VANET module and thus create a VANET and assign the roles of leading and following agents based on their available resources. 
	This process is repeated each time, a new vehicle comes in the VANET radius of the vehicles already in the VANET. 
	The resulting neighbors are added to the neighbors list in every agent and used for communication as the learning continues.
	The participating agents continuously compute the distance between each other using the values from their neighboring agent's sensors and maintain a safe distance to avoid collision.
	The DDPG module in Co-DDPG collects the necessary state information from the environment, VANET and collision avoidance modules and generates the optimal driving actions along with the associated reward under current conditions. 
	Using the parameter distribution module, the computed parameters from the leading agents are distributed to their following agents.
	Whenever a collision occurs, or if the vehicles overstep their safe distance threshold, a penalty with negative reward is awarded for that action.
	With a hierarchical networking architecture, RACE unites all the necessary autonomous driving components in Co-DDPG to provide an efficient learning platform that enhances the existing autonomous driving learning algorithms. 
	\begin{algorithm}[!t]
		\caption{checkVANETradius()}\label{alg:checkvanetradius}
		\begin{algorithmic}[1]
			\For{Neighbor in vehicle.Neighbor}
			\If{opponent $<$ VANET\_RADIUS}
			\If{not vehicle.openedVANET and not vehicle.insideVANET}
			\State{vehicle.openVANET()} 
			\Else
			\State{vehicle.identifyNeighbors()}
			\EndIf	
			\EndIf
			\EndFor
		\end{algorithmic}
	\end{algorithm}
	\setlength{\textfloatsep}{2pt}
	
	\section{VANET}\label{sec-vanet}
	In this section, we describe the dynamic vehicular ad-hoc network creation algorithm, \ie, the VANET component shown in Figure \ref{fi:framework}(b). 
	\vspace{-2mm}
	\subsection{Opening, Leaving and Closing VANETs}\label{subsec:openclose}
	The basic objective of a VANET during autonomous driving in RACE is achieved with three main functions: \texttt{openVANET()}, \texttt{leaveVANET()} and \texttt{closeVANET()}. 
	If a vehicle scanning for neighbors identifies another vehicle in its VANET range, then a VANET connection is opened using the \texttt{openVANET()} function to initialize the connection.
	Contrary to \texttt{openVANET()}, as soon as a vehicle moves out of the VANET's range, this vehicle leaves the network using \texttt{leaveVANET()} function and disconnects with its neighbors located in that network. 
	Subsequently, the neighbors of this vehicle remove its information from their list of connected vehicles. 
	When all except one vehicle leaves the VANET, then the scanning vehicle closes the VANET using the \texttt{closeVANET()} function. 
	
	\subsection{VANET Operation}\label{subsec:vehicleid}
	In order to create a VANET, it is fundamental to identify the vehicles driving in the area of interest. 
	Hence in RACE, we declare the desired VANETs range in meters and utilize the on-board sensors in the vehicles to compute the distance between vehicles located in this range.
	
	\begin{algorithm}[!t]
		\caption{identifyNeighbors()}\label{alg:identify}
		\begin{algorithmic}[1]
			\State{identifiedNeighbor = False}
			\For{Vehicle\emph{X} in LocalVehicleList}
			\For{Vehicle\emph{A} in LocalVehicleList}
			\If{ Vehicle\emph{X} != Vehicle\emph{A} }
			\If{Vehicle\emph{X} not in Vehicle\emph{A}.Neighbors \&\& Distance between \emph{A} and \emph{X} $<$ VANET\_RADIUS}
			\State{identifiedNeighbor = True}
			\State{Add Vehicle\emph{X} to  Vehicle\emph{A}.Neighbors}							\State{Add Vehicle\emph{A} to  Vehicle\emph{X}.Neighbors}
			\EndIf
			\EndIf	
			\EndFor
			\EndFor
		\end{algorithmic}
	\end{algorithm}
	
	In Co-DDPG, we define an algorithm called \texttt{checkVANET radius()} to find any neighboring vehicle in the specified VANET radius of a vehicle. 
	A pseudocode of the algorithm is given in Algorithm \ref{alg:checkvanetradius}.
	Essentially, every vehicle scans with \texttt{checkVANETradius()} the VANET radius \eg, 200 meters to find any nearby vehicles to connect with. 
	Since we follow a decentralized approach, every vehicle in RACE maintains a local list of connected vehicles called \texttt{LocalVehicleList}, which is a subset of the global list of all participating vehicles. 
	If a vehicle is found within this range and there is no available VANET to connect with, then the scanning vehicle initially opens a VANET using the \texttt{openVANET()} function.
	This is followed by opening a connection with the identified neighbor for communication.
	This is accomplished by calling the method \texttt{identifyNeighbors()} shown in Algorithm \ref{alg:identify}.
	All vehicles, continue to scan the VANET radius to find potential neighbors to connect with.
	Further, the \texttt{identifyNeighbors()} method is also responsible for maintaining the presently connected vehicles within a list.
	
	In addition to VANET creation, when a vehicle enters the VANET, it takes on the role of a leading or a following agent with each of its neighbors based on its available resources.
	Specifically, a new agent joining the VANET notifies its neighbours that it would like to be a leading or a following agent while its neighbours can choose to be either its leader or be a follower of the new agent.
	If a neighbouring agent is already a follower, then it notifies the new agent that it is already a following agent and provides the list of its leading agent(s).
	Since each vehicle available resources can vary, a following vehicle at level $i$ may become a leading agent for a vehicle at level $i+1$ if the resources of the agent at level $i+1$ are less than that of agents located at the level $i$.
	
\begin{figure}[!t]
		\centering
		\includegraphics[width=0.35\textwidth]{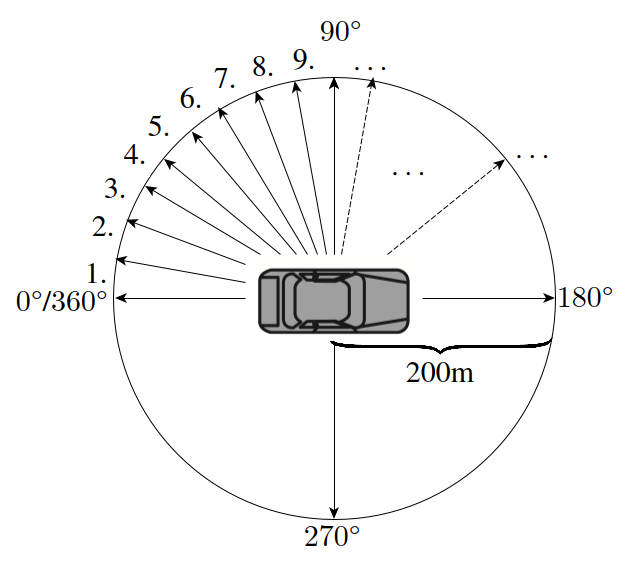}
		\vspace{-4mm}
		\caption{Distance computation in VANETs.}
		\label{fig:opponents}
	\end{figure}

	Utilizing the values gathered by the on-board sensors, the \texttt{identifyNeighbors()} method verifies the distance between the host vehicle and its neighbors and sends it to the collision avoidance component, which in turn sends this information to the Co-DDPG approach to learn collision avoidance during autonomous driving. 
	We implemented an autonomous driving system using the RACE framework in the TORCS simulator by primarily using the \emph{opponent range sensor}.
	According to the TORCS user manual \cite{loiacono2013simulated}, the opponent range measurement is achieved using 36 minor range on-board sensors located on each agent to cover the complete 360 degree space around the vehicle.
	A pictorial representation of the opponents range sensor in TORCS is shown in Figure \ref{fig:opponents}.
	Each of these sensors has a sensing range of up to 200 meters and span a coverage area of 10 degrees. 
	Each sensor returns the distance between the hosting vehicle and any nearby vehicle(s) within its sensing range. 
	
	\begin{algorithm}[!t]
		\caption{Co-DDPG algorithm}\label{alg:coddpg}
		\begin{algorithmic}[1]
			\State{Randomly initialize critic network $Q(s,a | \theta^Q)$ and actor  $\mu(s|\theta^\mu)$ with weights $\theta^Q$ and $\theta^\mu$}
			\State{Initialize target network $Q'$ and $\mu'$ with weights $\theta^{Q'} \leftarrow \theta^Q$, $\theta^{\mu'} \leftarrow \theta^\mu$}
			\State{Initialize action-value function $Q$ with random weights}
			\State{Initialize replay buffer $R$}
			\For{episode $= 1$ to $M$}
			\State{\textit{initializeVANET()}}
			\State{Initialize a random process $\mathcal{N}$ for action exploration}
			\State{Receive initial observation state $s_1$}
			\For{$t = 1$ to $T$}
			\State{\textit{setSensors($s_t$)}}
			\State{Select action $a_t = \mu(s_t | \theta^\mu) + \mathcal{N}_t$ according to the current policy and exploration noise}
			\State{Execute action $a_t$}
			\State{\textit{checkCollisionRadius()}}
			\State{\textit{Observe reward $r_t$ and observe new state $s_{t+1}$}}
			\State{\textit{checkVANETradius()}}
			\If{\textit{$t == 100$ and VANET is open}}
			\State{\textit{distributeParameters()}}
			\EndIf
			\State{Store transition $(s_t, a_t, r_t, s_{t+1})$ in $R$}
			\State{Sample a random minibatch of $N$ transitions $(s_i, a_i r_i, s_{i+1})$ from $R$}
			\State{Set $y_i = r_i + \gamma Q'(s_{i+1}, \mu'(s_{i+1} | \theta^{\mu'}) | \theta^{Q'})$}
			\State{Update critic by minimizing the loss: $L = \frac{1}{N}\sum_i (y_i - Q(s_i, a_i | \theta^Q))^2$}
			\State{Update the actor policy using the sampled policy gradient: $$\nabla_{\theta^\mu}J \approx \frac{1}{N}\sum_i \nabla_a Q(s,a | \theta^Q)|_{s = s_i, a = \mu(s_i)} \nabla_{\theta^\mu}\mu(s | \theta^\mu)|_{s_i}$$}
			\State{Update the target networks:
				\begin{align*}\theta^{Q'} \leftarrow \tau\theta^Q + (1 - \tau)\theta^{Q'} \\
				\theta^{\mu'} \leftarrow \tau\mu^Q + (1 - \tau)\mu^{Q'}
				\end{align*}}
			\EndFor
			\EndFor
		\end{algorithmic}
	\end{algorithm}
	\setlength{\textfloatsep}{2pt}
	\section{Co-DDPG}
	\label{sec-coddpg}
	In this section we describe the core of the proposed autonomous driving strategy designed in RACE, \ie, the system model for deep reinforcement learning in Co-DDPG. 
	
	Due to the compelling design of the standard DDPG \cite{ddpg}, we utilized it as the basis for learning in RACE. 
	DDPG can efficiently learn competitive policies for all tasks considered in an autonomous driving environment using low-dimensional observations and handle continuous action spaces effectively.
	However, we model the autonomous driving learning strategy through enhancing the standard DDPG algorithm with hierarchical multi-agent parameter distribution to further improve the performance of learning. 
	We refer to this enhanced version as Co-DDPG in this work and a pseudocode of the algorithm is provided in Algorithm \ref{alg:coddpg}.
	The actor network's update (line 23 in Algorithm \ref{alg:coddpg}) is based on the application of chain rules computed by the loss function (line 22 in Algorithm \ref{alg:coddpg}). 
	The target network's weights are then updated according to $\theta' \leftarrow \tau\theta + (1 - \tau)\theta'$, where $\tau << 1$. 
	The overall benefit in a multi-agent autonomous driving environment is achieved with the efficient collaboration of the surrounding components in the system design of Co-DDPG as shown in Figure \ref{fi:framework}(b). 
	Co-DDPG receives input from the VANET and Collision Avoidance modules to enforce efficient driving behavior.
	We further improve the learning by allotting positive rewards for good driving behaviour and penalties for bad driving behaviour with negative rewards.
	Thereby, the system model of Co-DDPG is created in RACE using three essential concepts, namely the \emph{state}, \emph{action} and \emph{reward}. 
	
	\subsection{State Space}
	The \emph{state} space represents the on-board sensors in a vehicle and is thus a representation of the autonomous driving environment.
	The \emph{state} is defined as $\mathcal{S}=\{s_{1}, s_{2}, \cdots, s_{I}\}$, where $I$ is the total number of states. 
	The states represent a vehicle's sensors and forms the basis for the actions taken by the agents.
	As mentioned before, we collected the state information from the sensors available in the TORCS simulator. 
	Please note that sensors provided in TORCS \eg, range finding sensors operate in a similar fashion as the real world sensors implemented with LiDAR, radar or ultrasonic sensor technologies.  
	Hence, the data produced by the sensors in TORCS enabled us to build a simulated learning environment akin to driving in the real world.
	Due to their prominent role, the sensors formed the basis for an autonomous vehicle to conduct any action in its environment.
	
	The following list of sensors, which are also the state space variables, are used to form the state space in Co-DDPG:
	\begin{itemize}
		\item \texttt{angle:} It is the angle between the vehicle axis and the track axis during driving with values  ranging between $-\pi$ to $\pi$ and is measured in rad.
		\item \texttt{opponents:} It is a vector containing the 36 range finding sensors from TORCS. Each opponent sensor spans a coverage area of 10 degrees with a sensing range of up to 200 meters. The vector returns the closest vehicle in the scanned area.
		\item \texttt{rpm:} It stands for the rotation per minute executed by the vehicle's engine.
		\item \texttt{v:} It denotes the vehicle's speed within the vehicle's longitudinal axis in km/h.
		\item \texttt{speedY:} It denotes the vehicle's speed within the vehicle's transverse axis in km/h.
		\item \texttt{speedZ:} It denotes the vehicle's speed within the vehicle's z-axis in km/h.
		\item \texttt{track:} It is a vector containing 19 range finding sensors that gather readings from the front of the vehicle with 90 degrees left and right from the vehicle's axis. Each sensor returns the distance between the vehicle and the track edge within a range of 200 meters.
		\item \texttt{trackPos:} It denotes the distance between the vehicle axis and the track axis. It returns a value of $0$ when the vehicle is found on the track axis, $-1$ when the vehicle is on the right track edge and $1$, when the vehicle is on the left track edge.
		\item \texttt{wheelSpinVel:} It represents a vector of four sensors capturing the rotation speed of the wheels.
	\end{itemize}
	\subsection{Action Space}
	The \emph{action} space is a three dimensional vector consisting of acceleration (i.e. accel), brake and steering. 
	It is denoted as $\mathcal{A}=\{a_{1}, a_{2},\cdots, a_{H}\}$, where $H$ is the total number of actions and is extracted from the following effectors:
	\begin{itemize}
		\item \texttt{accel:} It represents the simulated gas pedal and returns the values in the range of $[0,1]$, where $0$ represents no gas while $1$ represents full gas.
		\item \texttt{brake:} It represents the simulated brake pedal and returns values in the range of $[0,1]$, where $0$ represents no brake while $1$ represents full brake.
		\item \texttt{steering:} It represents the simulated steering values in the range of $[-1,1]$, where $-1$ represents full right brake while $1$ represents full left brake with an angle of $0.366519$ rad.
	\end{itemize}
	\begin{figure*}[!t]
		\centering
		\begin{minipage}{.49\textwidth}
			\centering
			\includegraphics[width=0.65\textwidth]{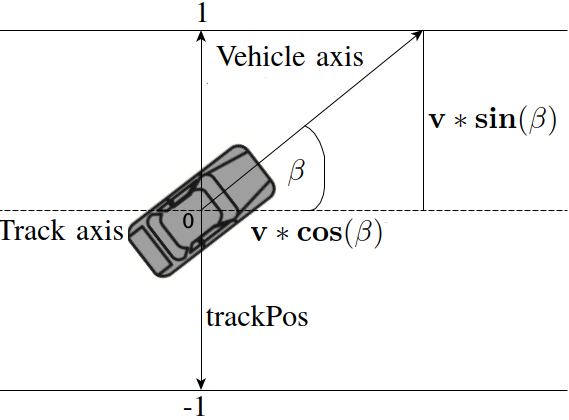}
			\vspace{-2mm}
			\caption{Reward computation.}
			\label{fig:rewardcomputation}
		\end{minipage}%
		\begin{minipage}{.49\textwidth}
			\centering
			\includegraphics[width=0.55\textwidth]{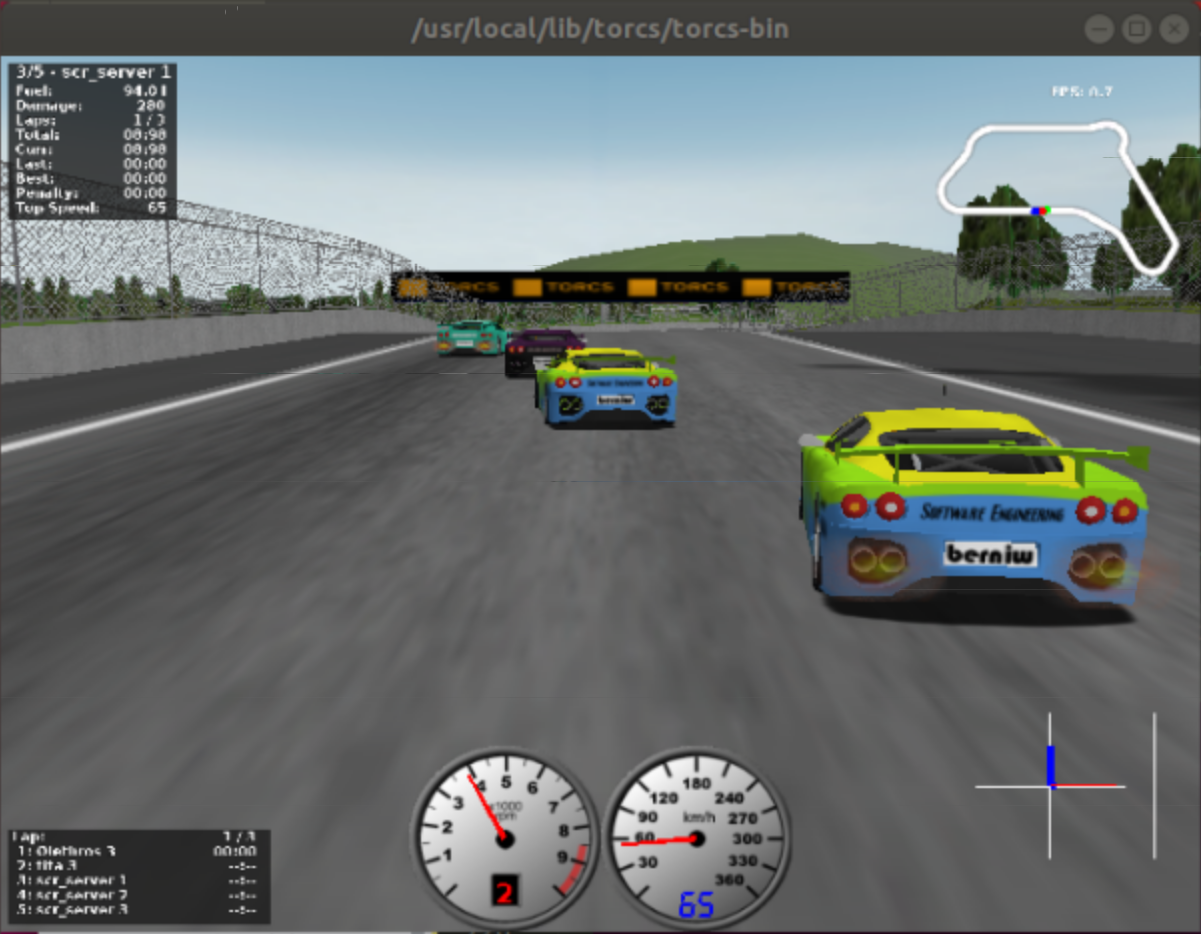}
			\caption{Experimental racing track.}
			\label{fig:cgtrack2}
		\end{minipage}%
		\vspace{-6mm}
	\end{figure*}
	
	\subsection{Reward}
	The reward function in Eq. \eqref{eq:reward}, which is our objective, is defined by considering three factors including collision avoidance, minimizing the arriving time of all vehicles and keeping vehicles on the road. 
	\begin{align}
	\label{eq:reward}
	r^{t}_{i}=w_{c}c^{t}_{i}+w_{h}h^{t}_{i}+w_{o}o^{t}_{i},
	\end{align}
	where $r^{t}_{i}$ denotes the reward $r$ of the $i$th vehicle at timestep $t$. Essentially, higher the reward value, better is the autonomous driving behavior.
	$c^{t}_{i}$, $h^{t}_{i}$ and $o^{t}_{i}$ are the rewards generated by avoiding collisions, minimizing the arriving time and keeping the vehicles on the road of the $i$th vehicle at timestep $t$, respectively. $w_{c}\in [0,1]$, $w_{h}\in [0,1]$ and $w_{o}\in [0,1]$ are the weights for them to define the contribution of each factor to the final reward function, where $w_{c}+w_{h}+w_{o}=1$.
	
	Co-DDPG obtains penalties by defining a $c^{t}_{i}$ to avoid collisions, and is given as follows,
	\begin{equation}
	c^{t}_{i} = 
	\begin{cases}
	-(D^{t}_{min,i}-d^{t}_{i})\cdot w_{p} & d^{t}_{i}\leq D^{t}_{min,i}, \\
	0 & \text{otherwise},
	\end{cases}
	\label{eq:collsionFactor}
	\end{equation}
	where $D^{t}_{min,i}$ calculated in Eq. \eqref{eq:minD} is the safe distance between two vehicles at timestep $t$, $w_p$ is the punished value per distance, $d^{t}_{i}$ is the distance between two vehicles at timestep $t$. For instance, for two vehicles $i$ and $j$, $d^{t}_{i}$ defined as $d^{t}_{i}=\sqrt{(x^t_i-x^t_j)^2+(y^t_i-y^t_j)^2}$. 
	Please note that we use the opponent sensors in the vehicles to get $d^{t}_{i}$ (See Section \ref{subsec:vehicleid}) and $D^{t}_{min,i}$ is the safe distance that should be maintained between vehicles. Similar to \cite{tientrakool2011highway}, we design the vehicle deceleration rate as $a^t_{cur} \in [a_{min}, a_{max}]$, whereas the safe distance between vehicles $v^{t}_{i}$ and $v^{v}_{j}$ is calculated as follows,
	\begin{equation}
	D^{t}_{min,i}=\frac{T_s v}{A}+\frac{v^{2}_t}{B|a^t_{0}|}-\frac{v^{2}_t}{B|a^t_{max}|},
	\label{eq:minD}
	\end{equation}
	where $T_{s}$ is the delay for detecting an emergency by on board sensors in the vehicles. This also incorporates the autonomous vehicle's brakes response time. We set $A=3.6$ to convert the vehicle speed $v$ from $km/h$ to $m/s$ while $B=25.92$.  B is defined as $2|a|\times (A)^2$ and it converts $v^2$ from $(km/h)^2$ to $(m/s)^2$.
	
	$h^{t}_{i}$ is the reward to minimize the traveling time, defined by
	\begin{equation}
	h^{t}_{i} = 
	\begin{cases}
	10w_{p} & \text{arrival to the destination}, \\
	w_{p}|v^{t}_{i} \times sin(\beta^{t}_{i})| & \text{otherwise}
	\end{cases}
	\label{eq:minimizeTime}
	\end{equation}
	where $w_p$ is the reward value as in Eq. \eqref{eq:collsionFactor}.
	
	$o^{t}_{i}$ is an incentive awarded to the vehicle to encourage good driving behaviour whereby the vehicle drives within the road, the factors of which are described in Figure \ref{fig:rewardcomputation}. It is calculated as follows,
	\begin{align}
	o^{t}_{i} = v^{t}_{i} \times |trackPos^{t}_{i}|  + \left(|v^{t}_{i} \times sin(\beta^{t}_{i})| -v^{t}_{i} \times cos(\beta^{t}_{i})\right),
	\label{eq:onRoad}
	\end{align}
	where $v^{t}_{i} \times cos(\beta^{t}_{i})$ depicts the value of the longitudinal velocity of vehicle $i$ at timestep $t$, which we want to minimize. 
	The term $|v^{t}_{i} \times sin(\beta^{t}_{i})|$ computes the transverse velocity and it is aimed to be as high as possible, since the vehicle should stay on the road. 
	Lower values indicate that the vehicle is going off the road edge. 
	The third term, $v^{t}_{i} \times |trackPos^{t}_{i}|$, also represents stable driving on the center of the road and hence it should be as high as possible. 
	In Co-DDPG, whenever $|v^{t}_{i} \times sin(\beta^{t}_{i})|$ and  $v^{t}_{i} \times |trackPos^{t}_{i}|$ are low, the reward is reduced by awarding penalties. 
	
	\section{Evaluation}
	In this section, we evaluate the benefits of RACE through demonstrating the performance of our system implemented using the RACE framework with the TORCS simulator. We further compare the performance of this system with existing baselines and show that RACE effectively outperforms the existing systems with respect to collisions and scalability; the pivotal requirements of autonomous driving. 
	\subsection{Assumptions}\label{subsec:assumptions}
	
	The following assumptions were made during the experiments to fulfill the basic requirements of autonomous driving.  
	Security and privacy are intrinsic components of the autonomous driving system, hence we ensure security by following the proposal from Samara \etal \cite{samara2017security}. 
	As for the privacy, RACE can guarantee the location privacy by using the relative distance to build VANET instead of the absolute location of the vehicles.
	
	We observed that underlying communication technology was necessary to ensure efficient operation of the overall system. Even though many communication technologies exist and newer technologies like 5G are gaining momentum \cite{cheng20175g,katsaros2019evolution}, the RACE framework is agnostic to these technologies. 
	Therefore, we made one assumption that any underlying communication technology that fulfills the requirements for performance, latency and parameter distribution in an autonomous driving environment could be used in RACE.
	
	Additionally, we made another assumption that VANETs created in the autonomous driving environment were robust in nature, wherein they effectively minimized connectivity failures and supported efficient data transfer among the participating agents. 
	This was necessary to maintain stable networks especially during long training/driving sessions.
	
	\subsection{TORCS Simulator}
	As mentioned earlier, we implemented a system using the RACE framework in the widely used \textit{TORCS}
	simulator developed by Wymann et al. \cite{wymann2015torcs}.
	Over the years, TORCS has become a profound testing environment for autonomous driving research \cite{sallab2017deep}. 
	Intrinsically, by including the basic vehicular dynamics and physics, TORCS provides realistic physical surroundings that are suitable for simulating desired driving environments. 
	The participating vehicles also known as players are characterized as \textit{robots} in TORCS and they are primarily external entities loaded by the simulation. 
	In particular, TORCS allows developers to create their own artificial intelligence agents to simulate desired experimental environments. 
	This is accomplished by a low-level API located at the driver level that enables partial access to the state of simulation including the status of the ongoing race or robot information such as agent's speed, agent's distance from the track edge or the distance to other participating agents. 
	Nevertheless, TORCS limits access to most of the components in the simulation and thus poses a challenge to even partially observe basic driving problems \cite{wymann2015torcs}. Despite such limitations, many TORCS robots were created over the years by researchers and developers to further examine the paradigms of Artificial Intelligence and Machine Learning. 
	Many of these research efforts have enhanced the TORCS system to a point that it is now possible to execute driving scenarios that surpasses human-level driving. 
	\subsection{Experimental Setup}\label{subsec:expsetup}
	We perform extensive experiments to measure three most important metrics of autonomous driving namely collisions, scalability and latency of the system during driving. 
	We use the racing environment with \textit{CG track 2} in TORCS shown in Figure \ref{fig:cgtrack2} as an example to illustrate the performance of RACE. Since CG track 2 is one typical track example in TORCS~\cite{cgtrack2}. The width and length of it are about \SI{3186}{m} and \SI{15}{m}, respectively. 
	
	We gradually introduced vehicles on to the racing track and observed their performance with respect to the effectiveness of their learning behaviour. For example, in these simulations, 10 Normal Driving Vehicles (NDVs) are running on the road. Then, we vary the number of Autonomous Driving Vehicles (ADVs) equipped with Co-DDPG to test the proposed RACE framework. We set weights of the reward function in Eq. \eqref{eq:reward} as $w_c=0.6$ and $w_h=w_o=0.2$. These values are obtained by comparing the performance from various simulations. The reward value of $w_p$ in Eq. \eqref{eq:collsionFactor} is set to 1000 and the communication range R of each vehicle is $R \in [0,200]$ meter. The speed range of vehicles is from \SI{40}{km/h} to \SI{60}{km/h}. Other parameters' values, like steering, acceleration, brake, angle, \etc, are listed in detail in the TORCS documentation \cite{loiacono2013simulated}. We implemented the RACE framework in TensorFlow where the Co-DDPG used two hidden layer neural networks as a non-linear function to achieve the optimal policy. There are 300 to 400 neurons in each layer, respectively. The learning rate is $1e-4$ with a batch size of 32.
	We use the decentralized sensor-level collision avoidance policy for multi-robot systems named POMDP proposed by Pinxin \etal \cite{long2018towards} as the baseline to compare  the performance of RACE. 
	POMDP uses a multi-stage deep reinforcement learning framework to help multiple robots to learn an optimal collision avoidance strategy using the policy gradient approach.
	
	One episode in the experiment denotes one instance on the real world racing track and thus one complete race from the start line until the finish line.
	However, there are instances when the race ends abruptly, such as when an agent turns back due to collision or leaves the track edge.
	As RACE is the foundation for the experiment, every leading agent trains with the Co-DDPG algorithm and shares the parameters with its following agents in the VANET. 
	One main vehicle is depicted as the leading agent that distributes its learned model parameters within one local VANET with its following agents and the following agents remain idle and await the learned parameters from their leading agent.
	Since we designed a non-deterministic autonomous driving environment, we executed each experiment ten times and computed the average results from all runs. 
	
	\subsection{RACE Performance Analysis}
	\label{subsec:racePerformance}
	Using the RACE system in TORCS, the agents learns the best driving behavior during the simulations to master collision avoidance policy. The agents' goal is to dynamically enhance their driving behaviour by learning to avoid collision with other agents and objects in their vicinity. Meanwhile, they also minimize their arrival time. 
	It is essential for an autonomous driving system to scale efficiently as the number of vehicles fluctuate. 
	Therefore, we measure the scalability of RACE under varying density of participating ADVs.
	We used ten NDVs along with one, two and three number of ADVs installed with Co-DDPG to measure the performance of RACE.
	
	The performance of ADVs is estimated by primarily examining the number of collisions experienced by the agents and the reward achieved by agents during the experiments, which is shown in Figure \ref{fi:ddpg10BoxReward} and \ref{fi:ddpg10BoxCollision}. The One CO-DDPG, Two Co-DDPG and Three Co-DDPG in the results denote ten NDVs with one, two and three ADVs operating with Co-DDPG in RACE, respectively.
	
	\begin{figure}
		\center
%
%
\definecolor{mycolor1}{rgb}{1.00000,0.00000,1.00000}%
\definecolor{mycolor2}{rgb}{0.00000,1.00000,1.00000}%
\begin{tikzpicture}

\begin{axis}[%
width=2.6052in,
height=1.7592in,
at={(0.758in,0.481in)},
scale only axis,
separate axis lines,
every outer x axis line/.append style={black},
every x tick label/.append style={font=\color{black}},
every x tick/.append style={black},
xlabel={Number of Episodes},
every outer y axis line/.append style={black},
every y tick label/.append style={font=\color{black}},
every y tick/.append style={black},
xmin=0,
xmax=500,
ymin=0,
ymax=70,
ytick={0, 20, 40, 60, 80, 100},
yticklabels={0, 20, 40, 60, 80, 100},
ylabel={Number of Collisions},
axis background/.style={fill=white},
xmajorgrids,
ymajorgrids,
legend style={at={(0.5,1.03)}, anchor=south, legend columns=1, legend cell align=left, align=left, draw=black, nodes={scale=0.7, transform shape}}
]
\addplot [color=black, dashed, line width=1.5pt, mark=triangle, mark options={solid, rotate=90, black}]
  table[row sep=crcr]{%
50	64\\
100	41\\
150	28\\
200	23\\
250	14\\
300 11\\
350 8\\
400 6\\
450 4\\
500 3\\
};
\addlegendentry{One Co-DDPG ADV with Ten NDVs in RACE}

\addplot [color=blue, dashed, line width=1.5pt, mark=o, mark options={solid, blue}]
  table[row sep=crcr]{%
50	40\\
100	29\\
150	18\\
200	13\\
250	9\\
300 7\\
350 4\\
400 0\\
450 0\\
500 0\\
};
\addlegendentry{Two Co-DDPG ADVs with Ten NDVs in RACE}

\addplot [color=red, dashed, line width=1.5pt, mark=o, mark options={solid, red}]
  table[row sep=crcr]{%
50	30\\
100	14\\
150	10\\
200	6\\
250	3\\
300 2\\
350 0\\
400 0\\
450 0\\
500 0\\
};
\addlegendentry{Three Co-DDPG ADVs with Ten NDVs in RACE}

\end{axis}
\end{tikzpicture}%
		\caption{The total number of collisions of ADVs over different number of episodes during the training process.}
		\label{fi:ddpg10BoxReward}
	\end{figure}
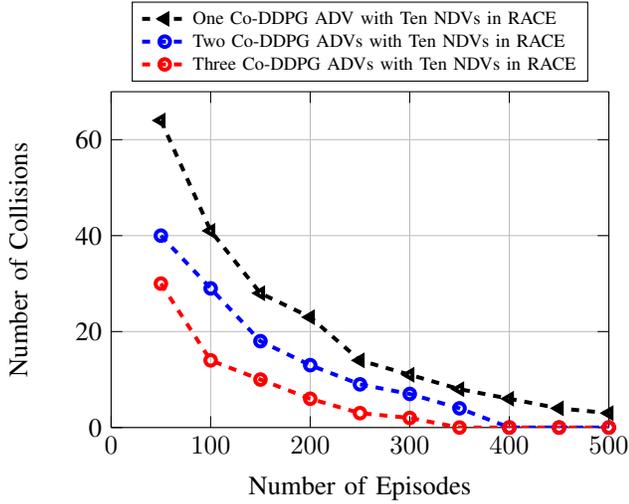
	\begin{figure}
		\center
%
%
\definecolor{mycolor1}{rgb}{1.00000,0.00000,1.00000}%
\definecolor{mycolor2}{rgb}{0.00000,1.00000,1.00000}%
\begin{tikzpicture}

\begin{axis}[%
width=2.6052in,
height=1.7592in,
at={(0.758in,0.481in)},
scale only axis,
separate axis lines,
every outer x axis line/.append style={black},
every x tick label/.append style={font=\color{black}},
every x tick/.append style={black},
xlabel={Number of Episodes},
every outer y axis line/.append style={black},
every y tick label/.append style={font=\color{black}},
every y tick/.append style={black},
xmin=0,
xmax=500,
ymin=-1000000,
ymax=1500000,
ylabel={Avg. Reward per ADV},
axis background/.style={fill=white},
xmajorgrids,
ymajorgrids,
legend style={at={(0.56,1.03)}, anchor=south, legend columns=1, legend cell align=left, align=left, draw=black, nodes={scale=0.7, transform shape}}
]
\addplot [color=black, dashed, line width=1.5pt, mark=triangle, mark options={solid, rotate=90, black}]
  table[row sep=crcr]{%
50	-615413\\
100	-314454\\
150	7843\\
200	57023\\
250	131075\\
300 199993\\
350 207617\\
400 243634\\
450 372085\\
500 392273\\
};
\addlegendentry{One Co-DDPG ADV with Ten NDVs in RACE}

\addplot [color=blue, dashed, line width=1.5pt, mark=o, mark options={solid, blue}]
  table[row sep=crcr]{%
50	-105940\\
100	218632\\
150	274722\\
200	380427\\
250	486519\\
300 571730\\
350 672859\\
400 775994\\
450 939399\\
500 986638\\
};
\addlegendentry{Two Co-DDPG ADVs with Ten NDVs in RACE}

\addplot [color=red, dashed, line width=1.5pt, mark=o, mark options={solid, red}]
  table[row sep=crcr]{%
50	52899\\
100	298971\\
150	424764\\
200	555820\\
250	673196\\
300 717951\\
350 886287\\
400 952018\\
450 1140163\\
500 1165089\\
};
\addlegendentry{Three Co-DDPG ADVs with Ten NDVs in RACE}

\end{axis}
\end{tikzpicture}%
		\caption{The average reward per ADV over different number of episodes during the training process.}
		\label{fi:ddpg10BoxCollision}
	\end{figure}
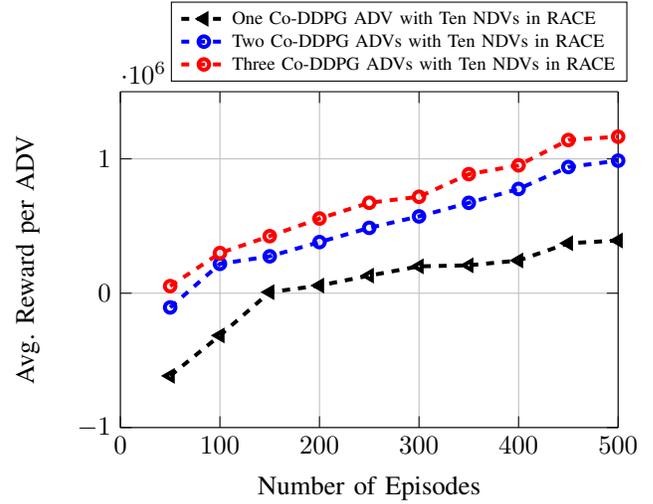
	
	In Figure \ref{fi:ddpg10BoxReward}, the average number of collisions over a span of 500 episodes during the training process is presented. The value of each collision point in this figure is calculated by summing the collision number within every 50 episodes.
	We observe from the results that as the number of episodes increased, the average number of collisions in all scenes are decreased. The number of collisions experienced by the Two Co-DDPG is between that of One Co-DDPG and Three Co-DDPG, as expected. Compared to Three Co-DDPG, the average number of collisions experiences by One Co-DDPG is around 211\% higher. This is because in RACE, the parameter distribution strategy (See Section \ref{subsec:parameterdist}) in Co-DDPG is employed to improve the performance of ADVs. We select the best driving policy for each vehicle by comparing the Co-DDPG rewards of neighboring autonomous driving vehicles in a VANET. The highest reward represents the best driving behaviours of the autonomous vehicles. Additionally, we also notice that longer training time is needed by One Co-DDPG to achieve zero collision in Figure \ref{fi:ddpg10BoxReward}. Whereas, in comparison with Two DDPG, the Three DDPG ADVs is 13\% faster in reaching the zero collision goal during training phases.
	
	The Co-DDPG rewards gain by the autonomous driving vehicles is measured and the results are shown in Figure \ref{fi:ddpg10BoxCollision}. The results describe the summation of the reward recorded during every 50 episodes. 
	We notice that as the number of episodes increases, the average reward increases in RACE. 
	In the beginning, the reward is low in all scenes, since Co-DDPG of each autonomous driving vehicle is initialized with random learning parameters in the initial stage. Therefore, the vehicles may not select the correct action for their next movement, which leads to many collisions. Co-DDPG punishes the failure actions by decreasing the rewards to help Co-DDPG learn from the bad behavior. Then, during next episode, it has the ability to choose the correct action based on its previous experiences. 
	As the learning experiences of the vehicles increase with further episodes, the rewards start to increase. This is because the vehicles learned to make a suitable action to avoid collisions. 
	In such scenarios, incentives are awarded for encouraging the desired driving behaviours and thereby minimize any collision.
	We observe that One Co-DDPG has the lowest reward over a span of 500 episodes compared to that in Two Co-DDPG and Three Co-DDPG. The reward per autonomous driving vehicle in Three Co-DDPG is about 33\% higher than that in Two Co-DDPG on average. 
	Moreover, our algorithm also awards penalties when there is a possibility for collision \eg, when the agents drive too close to each other.
	Interestingly, we observe that as they learned better policies, the agents choose to go off track rather than collide with other vehicles.
	
	\subsection{Comparison Analysis}
	In this section, we compare the performance of POMDP~\cite{long2018towards} with the proposed RACE framework. We utilize three metrics namely, collisions, reward and latency for this comparison. 
	
\begin{figure}[!t]
	\centering
	\includegraphics[width=0.5\textwidth]{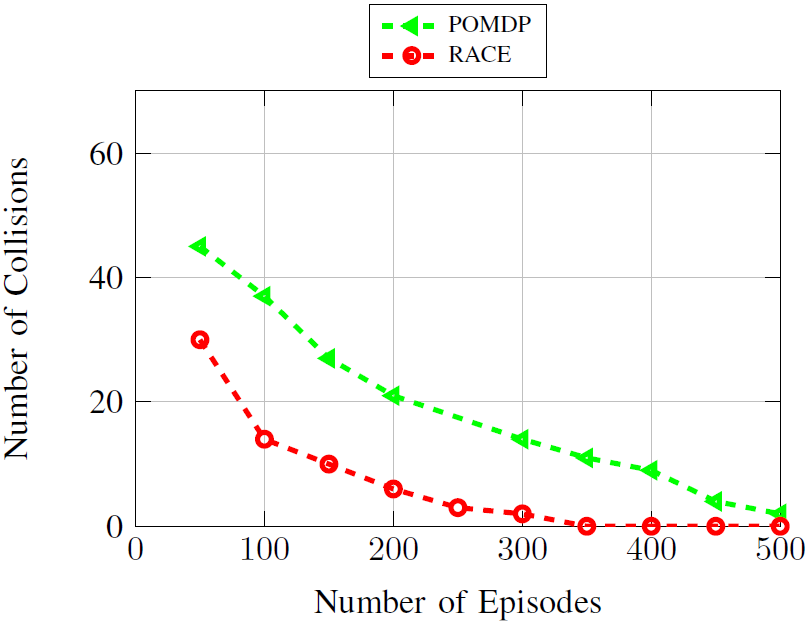}
	\vspace{-4mm}
	\caption{The comparison of total number of collisions of ADVs over the number of episodes in the training process.}
	\label{fi:ComCollision}
\end{figure}

	The comparison results showing the number of collisions in RACE and POMDP over different number of episodes are shown in Figure \ref{fi:ComCollision}, where three ADVs installed Co-DDPG and ten NDVs are used. Each value of point denotes a summation of the number of collisions experienced during every 50 episodes by all ADVs.
	We can see that number of collisions in RACE is around 65\% lower than that in POMDP on average during the training process. In addition, compared to POMDP, RACE learned the best collision avoidance policy by 42\% faster on average. 

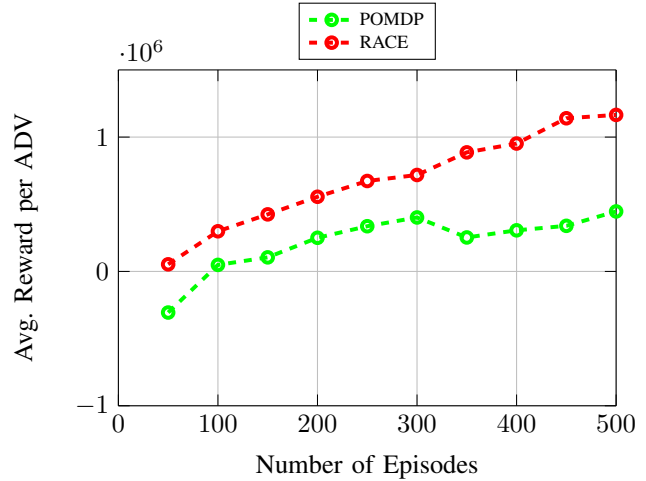
\begin{figure}
		\center
%
%
\definecolor{mycolor1}{rgb}{1.00000,0.00000,1.00000}%
\definecolor{mycolor2}{rgb}{0.00000,1.00000,1.00000}%
\begin{tikzpicture}

\begin{axis}[%
width=2.6052in,
height=1.7592in,
at={(0.758in,0.481in)},
scale only axis,
separate axis lines,
every outer x axis line/.append style={black},
every x tick label/.append style={font=\color{black}},
every x tick/.append style={black},
xlabel={Number of Episodes},
every outer y axis line/.append style={black},
every y tick label/.append style={font=\color{black}},
every y tick/.append style={black},
xmin=0,
xmax=500,
ymin=-1000000,
ymax=1500000,
ylabel={Avg. Reward per ADV},
axis background/.style={fill=white},
xmajorgrids,
ymajorgrids,
legend style={at={(0.5,1.03)}, anchor=south, legend columns=1, legend cell align=left, align=left, draw=black, nodes={scale=0.7, transform shape}}
]

\addplot [color=green, dashed, line width=1.5pt, mark=o, mark options={solid, green}]
  table[row sep=crcr]{%
50	-305940\\
100	48632\\
150	104722\\
200	250427\\
250	336519\\
300 401730\\
350 252859\\
400 305994\\
450 339399\\
500 446638\\
};
\addlegendentry{POMDP}

\addplot [color=red, dashed, line width=1.5pt, mark=o, mark options={solid, red}]
  table[row sep=crcr]{%
50	52899\\
100	298971\\
150	424764\\
200	555820\\
250	673196\\
300 717951\\
350 886287\\
400 952018\\
450 1140163\\
500 1165089\\
};
\addlegendentry{RACE}

\end{axis}
\end{tikzpicture}%
		\caption{The comparison of average reward per ADV over the number of episodes in the training process.}
		\label{fi:ComReward}
	\end{figure}

	Figure \ref{fi:ComReward} illustrates the average reward gained by every ADV in both POMDP and RACE. The simulation vehicles are same as in Figure \ref{fi:ComCollision}. 
	We add every 50 episodes' reward of all ADVs first. Then the summation reward value divides the number of ADVs to get the average reward point value in this figure.
	With increasing number of episodes, the average reward per ADV for both POMDP and RACE increases, which means that both of them can improve their driving policy by learning from the environment. 
	However, compared to POMDP, RACE achieved 215\% higher rewards. This demonstrates that RACE has the ability to learn better driving policies. 
	Because RACE not only uses the parameter distribution strategy to speed up the learning ability but also employs the adaptive dynamic reward function to process the collision scenario. 
	Besides, the learning process in RACE is more stable than in POMDP. 
	
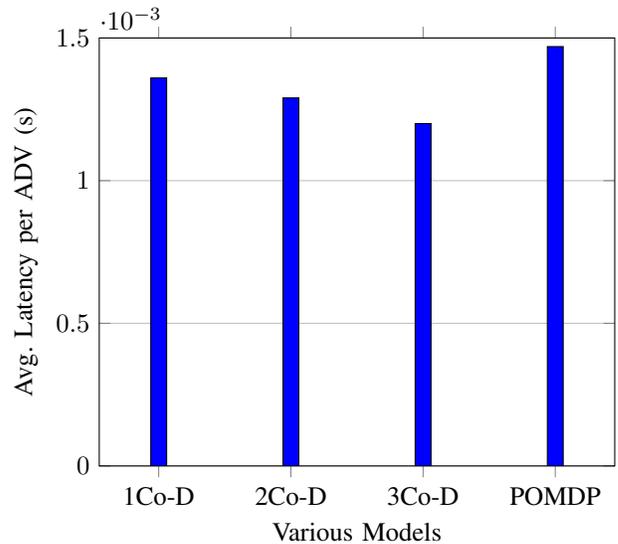
\begin{figure}
		\center
		    \begin{tikzpicture}
    \begin{axis}
    [
        symbolic x coords={1Co-D, 2Co-D, 3Co-D, POMDP},
        xtick=data,
        ylabel=Avg. Latency per ADV (s),
        every axis y label/.style={at={(ticklabel cs:0.5)},rotate=90,anchor=near ticklabel},
        ymin=0,ymax=0.0015,
        xlabel=Various Models,
        every axis x label/.style={at={(ticklabel cs:0.5)},anchor=near ticklabel},
        legend style={at={(0.5,-0.25)},
        anchor=north,legend columns=2},
        ybar,
        bar width=6pt,
        ymajorgrids = true,
        enlarge x limits=0.15
    ]
            \addplot[ybar,fill=blue] coordinates {
                (1Co-D,   0.00136)
                (2Co-D,  0.00129)
                (3Co-D,   0.00120)
                (POMDP,   0.00147)
            };
        \end{axis}
    \end{tikzpicture}
		\caption{The comparison of latency over different models in the testing process.}
		\label{fi:ComLatency}
	\end{figure}

	Figure \ref{fi:ComLatency} shows the latency per ADV over different models, where 1Co-D, 2Co-D and 3Co-D represents One Co-DDPG ADV with ten NDVs in RACE, Two Co-DDPG ADVs with ten NDVs in RACE and Three Co-DDPG ADVs with ten NDVs in RACE. 
	Latency is the time duration from the trained Co-DDPG in one ADV collecting states information from the environment to the time of obtaining its action, such as push the brake.
	We observed that 3Co-D in RACE has the lowest latency among all models. 2Co-D in RACE performed better than 1Co-D in RACE. 
	Since RACE utilizes the leading-following framework and parameter distribution strategy, a higher number of ADVs with Co-DDPG leads to learning better driving policies.
	Compared to POMDP, the latency per ADV was about 19\% lower in RACE with 3Co-D.  
	
	\section{Related Work}
	In this section we describe the related work through broadly classifying the relevant literature into two main categories, namely collision avoidance and multi-agent collision avoidance in an autonomous driving environment. 
	\subsection{Collision Avoidance}\label{sec:ca}
	Collision avoidance for autonomous vehicles and robots is an actively researched topic. 
	Various methods including rule-based \cite{lu2014rule}, optimization-based \cite{riegger2016centralized}, traditional machine learning \cite{tian2007dynamic,liu2005fuzzy} and deep learning methods \cite{long2018towards,long2017deep} have been proposed in recent years to minimize collisions during autonomous driving. 
	However, rule-based solutions do not easily adapt to different scenarios, while optimization-based methods may rapidly converge to local minimal solutions and hence, have relatively low computational efficiency. 
	In addition, a study by Namazi \etal \cite{namazi2019intelligent} shows that traditional machine learning-based solutions are not suitable for a complex and dynamic environment such as autonomous driving.
	Leveraging deep learning especially the Convolutional Neural Networks (CNNs), Lv \etal \cite{lv2015traffic} handled collision avoidance by predicting the traffic flow while Chen \etal \cite{chen2017decentralized} utilized DRL with multi-agents settings to avoid collisions.
	In addition, Cheng \etal \cite{cheng2019automated} formulated an automated enemy avoidance problem with Markov Decision Process and resolved it with temporal-difference reinforcement learning. 
	However, these proposals cannot effectively minimize the collisions since the participating robots do not communicate with each other. 
	Chen \etal \cite{chen2018rear} predicted rear-end collisions using GA-optimized neural networks. Collisions between cyclists and heavy goods vehicles were prevented in \cite{jia2016field} by solving the side-to-side collisions problem. However, these approaches did not consider collisions from any other directions.
	On the other hand, Long \etal \cite{long2018towards} used deep learning to avoid collisions among robots and Lu \etal \cite{lu2018deep} investigated interaction between three pedestrians and one robot while Sangiovanni \etal \cite{sangiovanni2018deep} considered an unpredictable object as an obstacle to avoid collision. 
	All of the above mentioned proposals failed to consider dynamic and complex environments with possibly high speed vehicles. 
	\subsection{Multi-Agent Collision Avoidance}\label{sec:maenv}
	In the following we discuss the recent proposals that avoid collisions in a multi-agent environment.
	
	In \cite{deng2019cooperative,gelbal2019cooperative}, authors utilize connected networks and simulated overtaking maneuvers to avoid collisions.
	However, these proposals mainly focus on a specific type of collision and use complex mathematical models instead of machine learning methods. 
	Hence they require expert knowledge to design efficient mathematical models. 
	In addition, long \etal \cite{long2017deep} proposed sensor-level collision avoidance policies. 
	However, the supervised policies face limitations during learning w.r.t training dataset.
	In \cite{lou2018formation}, Lou \etal proposed a formation control law using a combination of consensus-based formation control method and a collision avoidance algorithm, 
	while lu \etal \cite{lu2018deep} proposed a human-aware algorithm to find a smooth and collision-free path. 
	However, these algorithms are computationally intensive and were not tested in complex environments and hence their effectiveness is difficult to assess.
	
	In \cite{foerster2016learning,dentler2019collision}, authors proposed centralized solutions.
	However, centralized learning produces obvious limitations for effective transmissions of the policies.
	The high computational burden of an optimization-based centralized scheme makes the deployment of the control system on real platforms challenging. On the other hand, Chen \etal \cite{chen2017decentralized} developed a decentralized multi-agent collision avoidance algorithm where two agents were simulated to navigate toward their own goal positions and learn a value network that encodes the expected time to goal. 
	However, cooperative information among robots is not accounted for in the solution and the design is not suitable for high speed scenarios.
	The Optimal Reciprocal Collision Avoidance (ORCA) framework from Van \etal \cite{van2011reciprocal} along with its extensions  \cite{snape2011hybrid,bareiss2015generalized} is widely used in crowd simulation and multi-agent systems. 
	It provides the necessary conditions for multiple robots to avoid collisions with each other in a short time span, and easily scales to handle large systems with many robots. 
	However, these methods are sensitive to the uncertainties in the real world driving scenarios as they assume that each robot has accurate information about the surrounding agents' positions, velocities and shapes.
	
	Based on the above mentioned literature and our extensive studies, we believe that RACE provides a more private, dynamic and efficient solution for collision avoidance during autonomous driving.
	In comparison to existing approaches, RACE is more dynamic than rule-based methods and has less computational complexity than optimization-based and other above mentioned deep leaning methods. 
	Hence, RACE optimally enhances autonomous driving behaviors of vehicles and effectively minimizes collisions during driving.

	\section{Conclusion}
	In this work, we studied the requirements of autonomous driving in detail and provided the necessary design goals.
	We proposed an efficient and decentralized framework called RACE to fulfill these goals.
	We developed an efficient collision avoidance algorithm named Co-DDPG which exploits a hierarchical multi-agent environment in VANET. The rewards in Co-DDPG are defined by considering the collision avoidance, minimizing arrival time and maintaining vehicles on the road. Its value measures the quality of driving behavior e.g., a higher reward represents a better driving behaviour.  
	The parameter distribution strategy is employed to expedite the learning process.
	Using a location-private communication strategy with a leader-follower multi-agent architecture, we distribute the best policies in the VANET and conserve important resources during driving.
	With extensive experiments in TORCS, we showed that RACE effectively reduced the number of collisions and scaled effortlessly with increasing number of autonomous vehicles. 
	We also demonstrated the meticulous learning in RACE, where vehicles learned to go off-track instead of colliding with other vehicles. 
	This improved the driving behaviour and minimized the associated risk and liability.
	As part of future work, we plan to enhance RACE through incorporating images or videos of the surroundings by camera during driving into the learning process. Additionally, we consider to improve RACE to adapt to the scenario, where the road has no speed range requirement.
	\section*{Acknowledgement}
	This project has received funding from the European Union's Horizon 2020 COSAFE project under grant agreement No 824019 and the China Scholarship Council under Grant ID 201706050095.
	\ifCLASSOPTIONcaptionsoff
	\newpage
	\fi
	\bibliographystyle{ieeetr}
	\bibliography{refs}
\end{document}